\documentclass{article}%
\usepackage{amsfonts}
\usepackage{amsmath}
\usepackage{amssymb}
\usepackage{graphicx}%
\providecommand{\U}[1]{\protect\rule{.1in}{.1in}}

\begin{document}

\title{N=1 Supergravity and Maxwell superalgebras}
\author{P. K. Concha$^{1,2,3}$, E. K. Rodr\'{\i}guez$^{1,2,3}$\\$^{1}${\small {\textit{Departamento de F\'{\i}sica, Universidad de
Concepci\'{o}n,}} }\\{\small {\textit{Casilla 160-C, Concepci\'{o}n, Chile.}} }\\\ $^{2}${\small {\textit{Dipartimento di Scienza Applicata e Tecnologia
(DISAT),}} }\\{\small {\textit{Politecnico di Torino, Corso Duca degli Abruzzi 24,}}}\\{\small {\textit{I-10129 Torino, Italia.}} }\\\ $^{3}${\small {\textit{Istituto Nazionale di Fisica Nucleare (INFN) Sezione
di Torino,}} }\\{\small {\textit{Via Pietro Giuria 1, 10125 Torino, Italia.}}}\\
\\{\small {\textit{E-mail:} {\texttt{patrickconcha@udec.cl},
\texttt{everodriguez@udec.cl}}}}}
\maketitle

\begin{abstract}
We present the construction of the $D=4$ supergravity action from the minimal
Maxwell superalgebra $s\mathcal{M}_{4}$, which can be derived from the
$\mathfrak{osp}\left(  4|1\right)  $ superalgebra by applying the abelian
semigroup expansion procedure. We show that $N=1$, $D=4$ pure supergravity can
be obtained alternatively as the MacDowell-Mansouri like action built from the
curvatures of the Maxwell superalgebra $s\mathcal{M}_{4}$. \ We extend this
result to all minimal Maxwell superalgebras type $s\mathcal{M}_{m+2}$. The
invariance under supersymmetry transformations is also analized.

\end{abstract}

\section{Introduction}

\qquad It is well known that the so-called Maxwell algebra $\mathcal{M}$
corresponds to a modification of the Poincar\'{e} symmetries, where a constant
electromagnetic field background is added to the Minkowski space \cite{1, 2,
3, 4, 5, 6}. \ In $D=4$ this algebra is obtained by adding to the Poincar\'{e}
generators $\left(  J_{ab},P_{a}\right)  $ the tensorial central charges
$Z_{ab}$, modifying the commutativity of the translation generators $P_{a}$ as
follows%
\begin{equation}
\left[  P_{a},P_{b}\right]  =Z_{ab}.
\end{equation}
In this way, the Maxwell algebra is an enlargement of Poincar\'{e} algebra,
i.e., if we consider $Z_{ab}=0$ we recover the Poincar\'{e} algebra.

Recently, it was shown that the Maxwell algebra can be obtained as an
expansion procedure of the $AdS$ Lie algebra $\mathfrak{so}(3,2)$ \cite{7, 8}.
\ In particular, in Ref. \cite{8} it was shown that the Maxwell algebra can be
derived using the $S$-expansion procedure introduced in Ref. \cite{9}, using
$S_{E}^{\left(  2\right)  }=\{\lambda_{0},\lambda_{1},\lambda_{2},\lambda
_{3}\}$ as the relevant semigroup . \ Furthermore, this result was extended to
all Maxwell algebras type $\mathcal{M}_{m}$ which can be obtained as an
$S$-expansion of the $AdS$ algebra using $S_{E}^{\left(  N\right)  }=\left\{
\lambda_{\alpha}\right\}  _{\alpha=0}^{N+1}$ as an abelian semigroup
\cite{10}. \ The $S$-expansion procedure is not only an useful method to
derive new Lie (super)algebras but it is a powerful tool in order to build new
(super)gravity theories. \ For example, it was also shown that standard
odd-dimensional General Relativity can be obtained from Chern--Simons gravity
theory for these $\mathcal{M}_{m}$ algebras%
\footnote{Also known as generalized Poincar\'{e}
algebras
$\mathfrak{B}_{m}$.}
and recently it was found that standard even-dimensional General Relativity
emerges as a limit of a Born-Infeld like theory invariant under a certain
subalgebra of the Lie algebra $\mathcal{M}_{m}$ \cite{10, 11, 12, 13}.

In Ref. \cite{14}, it was shown that the $N=1,$ $D=4$ Maxwell superalgebra
$s\mathcal{M}$ can be obtained as an enlargement of the Poincar\'{e}
superalgebra. \ This is particularly interesting since it describes the
supersymmetries of generalized $N=1$, $D=4$ superspace in the presence of a
constant abelian supersymmetric field strength background. \ Very recently it
was shown that minimal Maxwell superalgebra $s\mathcal{M}\ $can be obtained
using the Maurer Cartan expansion method \cite{7}. $\ $Subsequently, this
superalgebra and its generalization $s\mathcal{M}_{m+2}$ have been obtained as
an $S$-expansion of the $\mathfrak{osp}\left(  4|1\right)  $ superalgebra
\cite{15}. \ This family of superalgebras which contain the Maxwell algebras
type $\mathcal{M}_{m+2}$ as bosonic subalgebras may be viewed as a
generalization of the D'Auria-Fr\'{e} superalgebra \cite{16} and Green
algebras \cite{17}.

It is the purpose of this work to construct the minimal $D=4$ supergravity
action from the minimal Maxwell superalgebra $s\mathcal{M}_{4}$. \ To this
aim, we apply the $S$-expansion procedure to the $\mathfrak{osp}\left(
4|1\right)  $ superalgebra and we build a Mac Dowell-Mansouri like action with
the expanded $2$-form curvatures. \ We show that $N=1$, $D=4$ pure
supergravity can be derived alternatively as the MacDowell-Mansouri like
action from the minimal Maxwell superalgebra $s\mathcal{M}_{4}$. This result
corresponds to a supersymmetric extension of Ref. \cite{12} in which
four-dimensional General Relativity is derived from Maxwell algebra as a
Born-Infeld like action.\ \ We extend this result to all minimal Maxwell
superalgebras type $s\mathcal{M}_{m+2}$ in $D=4$. Interestingly, when the
simplest Maxwell superalgebra is considered we obtain the action found in Ref.
\cite{18}.

This work is organized as follows: In Section II, we briefly review the
principal aspects of the $S$-expansion procedure. \ In section III, we review
the $N=1$, $D=4$ supergravity with cosmological constant for the
$\mathfrak{osp}\left(  4|1\right)  $ superalgebra. \ Section IV and V contain our
main results. \ In section IV, we obtain the supergravity action as a Mac
Dowell-Mansouri like action from the Maxwell superalgebra $s\mathcal{M}_{4}$.
\ \ We show that this action describes pure supergravity. \ In section V we
extend our results to all minimal Maxwell superalgebras type $s\mathcal{M}%
_{m+2}$ and we study the invariance under supersymmetry. \ Section VI
concludes the work with some comments.

\section{S-expansion procedure}

\qquad It is the purpose of this section to review the main properties of the
$S$-expansion method introduced in Ref. \cite{9}.

The $S$-expansion procedure consists in combining the inner multiplication law
of a semigroup $S$ with the structure constants of a Lie algebra
$\mathfrak{g}$. Let $S=\left\{  \lambda_{\alpha}\right\}  $ be an abelian
semigroup with 2-selector $K_{\alpha\beta}^{\ \ \ \gamma}$ defined by
\begin{equation}
K_{\alpha\beta}^{\ \ \ \gamma}=\left\{
\begin{array}
[c]{cc}%
1 & \ \ \ \ \ \ \ \ \ \ \mbox{when}\ \ \lambda_{\alpha}\lambda_{\beta}%
=\lambda_{\gamma},\\
0 & \mbox{otherwise},
\end{array}
\right.
\end{equation}
and $\mathfrak{g}$ a Lie (super)algebra with basis $\left\{  \mathbf{T}%
_{A}\right\}  $ and structure constants $C_{AB}^{\ \ \ C}$,
\begin{equation}
\left[  \mathbf{T}_{A},\mathbf{T}_{B}\right]  =C_{AB}^{\ \ \ C}\mathbf{T}%
_{C}\text{.}%
\end{equation}
Then, the direct product $\mathfrak{G}=S\times\mathfrak{g}$ is also a Lie
(super)algebra with structure constants $C_{(A,\alpha)(B,\beta)}%
^{\ \ \ \ \ \ \ \ \ \ \ \ (C,\gamma)}=K_{\alpha\beta}^{\ \ \gamma}%
C_{AB}^{\ \ \ \ C}$, given by
\begin{equation}
\left[  \mathbf{T}_{(A,\alpha)},\mathbf{T}_{(B,\beta)}\right]  =C_{(A,\alpha
)(B,\beta)}^{\ \ \ \ \ \ \ \ \ \ \ \ (C,\gamma)}\mathbf{T}_{(C,\gamma)}.
\end{equation}
The Lie algebra $\mathfrak{G}$ defined by $\mathfrak{G}=S\times\mathfrak{g}$
is called $S$-expanded algebra of $\mathfrak{g}$.

When the semigroup has a zero element $0_{S}\in S$, it plays a somewhat
peculiar role in the $S$-expanded algebra. The algebra obtained by imposing
the condition $0_{S}\mathbf{T}_{A}=0$ on $\mathfrak{G}$ is called $0_{S}%
$-reduced algebra of $\mathfrak{G}$.

Interestingly, it is possible to extract smaller algebras from $S\times
\mathfrak{g}$. \ However, before to extract smaller algebras it is necessary
to apply a decomposition of the original algebra $\mathfrak{g}$. \ Let
$\mathfrak{g}=\bigoplus_{p\in I}V_{p}$ be a decomposition of $\mathfrak{g}$ in
subspaces $V_{p}$, where $I$ is a set of indices. \ Then for each $p,q\in I$
it is always possible to define $i_{\left(  p,q\right)  }\subset I$ such that%
\begin{equation}
\left[  V_{p},V_{q}\right]  \subset\bigoplus\limits_{r\in i_{\left(
p,q\right)  }}V_{r}. \label{eq33}%
\end{equation}
\textbf{\ }Now, let $S=\bigcup_{p\in I}S_{p}$ be a subset decomposition of the
abelian semigroup $S$ such that%
\begin{equation}
S_{p}\cdot S_{q}\subset\bigcup_{r\in i_{\left(  p,q\right)  }}S_{p}.
\label{eq34}%
\end{equation}
When such subset decomposition exists, then we say that this decomposition is
in resonance with the subspace decomposition of $\mathfrak{g}$. Defining the
subspaces of $\mathfrak{G}=S\times\mathfrak{g}$,%
\begin{equation}
W_{p}=S_{p}\times V_{p},\text{ \ }p\in I
\end{equation}
we have that
\begin{equation}
\mathfrak{G}_{R}=\bigoplus_{p\in I}W_{p},
\end{equation}
is a subalgebra of $\mathfrak{G}=S\times\mathfrak{g}$, and it is called a
resonant subalgebra of the $S$-expanded algebra $\mathfrak{G}$.

Another case of smaller algebra can be derived when the semigroup has a zero
element $0_{S}\in S$. The algebra obtained by imposing the condition
$0_{S}\mathbf{T}_{A}=0$ on $\mathfrak{G}$ is called $0_{S}$-reduced algebra of
$\mathfrak{G}$. \ Additionally a reduced algebra can be extracted from a
resonant subalgebra.

A useful property of the $S$-expansion procedure is that it provides us with
an invariant tensor for the $S$-expanded algebra $\mathfrak{G}=S\times
\mathfrak{g}$ in terms of an invariant tensor for $\mathfrak{g}$. As was shown
in Ref. \cite{9} the theorem VII.1 provide a general expression for an
invariant tensor for an expanded algebra.

\textbf{Theorem VII.1:} \ Let $S$ be an abelian semigroup, $\mathfrak{g}$ a
Lie (super)algebra of basis $\left\{  \mathbf{T}_{A}\right\}  $, and let
$\langle\mathbf{T}_{A_{n}}\cdots\mathbf{T}_{A_{n}}\rangle$ be an invariant
tensor for $\mathfrak{g}$. Then, the expression%
\begin{equation}
\langle\mathbf{T}_{(A_{1},\alpha_{1})}\cdots\mathbf{T}_{(A_{n},\alpha_{n}%
)}\rangle=\alpha_{\gamma}K_{\alpha_{1}\cdots\alpha_{n}}%
^{\ \ \ \ \ \text{\ \ \ \ }\gamma}\langle\mathbf{T}_{A_{1}}\cdots
\mathbf{T}_{A_{n}}\rangle
\end{equation}
where $\alpha_{\gamma}$ are arbitrary constants and $K_{\alpha_{1}\cdots
\alpha_{n}}^{\ \ \ \ \ \text{\ \ \ \ }\gamma}$ is the $n$-selector for $S$,
corresponds to an invariant tensor for the $S$-expanded algebra $\mathfrak{G}%
=S\times\mathfrak{g}$.

The proofs of these definitions and theorem can be found in Ref. \cite{9}.

\section{$N=1$, $D=4$ AdS Supergravity}

\qquad In Ref. \cite{19} was presented a geometric formulation of $N=1$
supergravity in four dimensions, where the relevant gauge fields of the theory
are those corresponding to the $\mathfrak{osp}(4|1)$ supergroup. The resulting
action, constructed only in terms of the gauge fields, leads to $N=1$
supergravity plus cosmological and topological terms. \ In this section a
brief review of this construction is considered.

The (anti)-commutation relations for the $\mathfrak{osp}\left(  4|1\right)  $
superalgebra are given by%
\begin{align}
\left[  \tilde{J}_{ab},\tilde{J}_{cd}\right]   &  =\eta_{bc}\tilde{J}%
_{ad}-\eta_{ac}\tilde{J}_{bd}-\eta_{bd}\tilde{J}_{ac}+\eta_{ad}\tilde{J}%
_{bc},\label{ADS01}\\
\left[  \tilde{J}_{ab},\tilde{P}_{c}\right]   &  =\eta_{bc}\tilde{P}_{a}%
-\eta_{ac}\tilde{P}_{b},\\
\left[  \tilde{P}_{a},\tilde{P}_{b}\right]   &  =\tilde{J}_{ab}%
,\label{AdSalgebra}\\
\left[  \tilde{J}_{ab},\tilde{Q}_{\alpha}\right]   &  =-\frac{1}{2}\left(
\gamma_{ab}\tilde{Q}\right)  _{\alpha},\text{ \ \ \ \ }\left[  \tilde{P}%
_{a},\tilde{Q}_{\alpha}\right]  =-\frac{1}{2}\left(  \gamma_{a}\tilde
{Q}\right)  _{\alpha},\\
\left\{  \tilde{Q}_{\alpha},\tilde{Q}_{\beta}\right\}   &  =-\frac{1}%
{2}\left[  \left(  \gamma^{ab}C\right)  _{\alpha\beta}\tilde{J}_{ab}-2\left(
\gamma^{a}C\right)  _{\alpha\beta}\tilde{P}_{a}\right]  , \label{ADS05}%
\end{align}
where $\tilde{J}_{ab}$, $\tilde{P}_{a}$ and $\tilde{Q}_{\alpha}$ correspond to
the Lorentz generators, the $AdS$ boost generators and the fermionic
generators, respectively.

\ In order to write down a Lagrangian for this algebra, we start from the
one-form gauge connection%
\begin{equation}
A=\frac{1}{2}\omega^{ab}J_{ab}+\frac{1}{l}e^{a}P_{a}+\frac{1}{\sqrt{l}}%
\psi^{\alpha}Q_{\alpha},
\end{equation}
and the associated two-form curvature $F=dA+A\wedge A$%
\begin{equation}
F=F^{A}T_{A}=\frac{1}{2}\mathcal{R}^{ab}J_{ab}+\frac{1}{l}R^{a}P_{a}+\frac
{1}{\sqrt{l}}\rho^{\alpha}Q_{\alpha}, \label{Curva}%
\end{equation}
where%
\begin{align*}
\mathcal{R}^{ab}  &  =d\omega^{ab}+\omega_{\text{ }c}^{a}\omega^{cb}+\frac
{1}{l^{2}}e^{a}e^{b}+\frac{1}{2l}\bar{\psi}\gamma^{ab}\psi,\\
R^{a}  &  =de^{a}+\omega_{\text{ }b}^{a}e^{b}-\frac{1}{2}\bar{\psi}\gamma
^{a}\psi,\\
\rho &  =d\psi+\frac{1}{4}\omega_{ab}\gamma^{ab}\psi+\frac{1}{2l}e^{a}%
\gamma_{a}\psi=D\psi+\frac{1}{2l}e^{a}\gamma_{a}\psi.
\end{align*}
The one-forms $e^{a},\omega^{ab}$ and $\psi$ are respectively the vierbein,
the spin connection and the gravitino field (a Majorana spinor, i.e,
$\bar{\psi}=\psi^{T}C$, where $C$ is the charge conjugation matrix).

Here we have introduced a length scale $l$. \ This is done because we have
chosen the Lie algebra generators $T_{A}=\left\{  J_{ab},P_{a},Q_{\alpha
}\right\}  $ as dimensionless and thus the one form connection $A=A_{\text{
}\mu}^{A}T_{A}dx^{\mu}$ must also be dimensionless. \ However, the vierbein
$e^{a}=e_{\text{ }\mu}^{a}dx^{\mu}$ must have dimensions of length if it is
related to the spacetime metric $g_{\mu\nu}$ through the usual equation
$g_{\mu\nu}=e_{\text{ }\mu}^{a}e_{\text{ }\nu}^{b}\eta_{ab}$. \ This means
that the "true" gauge field must be considered as $e^{a}/l$, with $l$ a length
parameter. In the same way, as the gravitino $\psi=\psi_{\mu}dx^{\mu}$ has
dimensions of $\left(  \text{length}\right)  ^{1/2}$, we must consider that
$\psi/\sqrt{l}$ is the gauge field of supersymmetry.

The general form of an action constructed with the 2-form curvature $\left(
\ref{Curva}\right)  $ is given by
\begin{equation}
S=2\int\left\langle F\wedge F\right\rangle =2\int F^{A}\wedge F^{B}%
\left\langle T_{A}T_{B}\right\rangle . \label{actmm}%
\end{equation}
Let us note that if we choose $\left\langle T_{A}T_{B}\right\rangle $ as an
invariant tensor\ (which satisfies the Bianchi identity) for the $Osp\left(
4|1\right)  $ supergroup, then the action $\left(  \ref{actmm}\right)  $ is a
topological invariant and gives no equations of motion. \ Nevertheless, with
the following choice of the invariant tensor
\begin{equation}
\left\langle T_{A}T_{B}\right\rangle =\left\{
\begin{array}
[c]{l}%
\left\langle J_{ab}J_{cd}\right\rangle =\epsilon_{abcd}\\
\left\langle Q_{\alpha}Q_{\beta}\right\rangle =2\left(  \gamma_{5}\right)
_{\alpha\beta}%
\end{array}
\right.  \label{Tinvads}%
\end{equation}
the action $\left(  \ref{actmm}\right)  $ becomes%
\begin{equation}
S=2\int\frac{1}{4}\mathcal{R}^{ab}\mathcal{R}^{ab}\epsilon_{abcd}+\frac{2}%
{l}\bar{\rho}\gamma_{5}\rho
\end{equation}
which corresponds to the Mac Dowell-Mansouri action \cite{19}. \ This choice
of the invariant tensor, which is necessary in order to reproduce a dynamical
action, breaks the $Osp\left(  4|1\right)  $ supergroup to their Lorentz subgroup.

The explicit form of the action is given by,
\begin{align}
S  &  =\int\frac{1}{2}\epsilon_{abcd}\left(  R^{ab}R^{cd}+\frac{2}{l^{2}%
}R^{ab}e^{c}e^{d}+\frac{1}{l^{4}}e^{a}e^{b}e^{c}e^{d}+\frac{2}{l^{3}}\bar
{\psi}\gamma^{ab}\psi e^{c}e^{d}\right) \nonumber\\
&  +\frac{4}{l^{2}}\bar{\psi}e^{a}\gamma_{a}\gamma_{5}D\psi+\frac{4}%
{l}d\left(  \bar{\psi}\gamma_{5}D\psi\right)
\end{align}
which can be written, modulo boundary terms, as follow$\ $%
\begin{equation}
S=\int\frac{1}{l^{2}}\left(  \epsilon_{abcd}R^{ab}e^{c}e^{d}+4\bar{\psi}%
e^{a}\gamma_{a}\gamma_{5}D\psi\right)  +\frac{1}{2}\epsilon_{abcd}\left(
\frac{1}{l^{4}}e^{a}e^{b}e^{c}e^{d}+\frac{2}{l^{3}}\bar{\psi}\gamma^{ab}\psi
e^{c}e^{d}\right)  . \label{adsaction}%
\end{equation}
The action $\left(  \ref{adsaction}\right)  $ corresponds to the Mac
Dowell-Mansouri action for the $\mathfrak{osp}\left(  4|1\right)  $
superalgebra \cite{19, 20}. This action, describing $N=1$, $D=4$ supergravity,
is not invariant under the $\mathfrak{osp}\left(  4|1\right)  $ gauge
transformations. \ However the invariance of the action under supersymmetry
transformation can be obtained modifying the spin connection $\omega^{ab}$
supersymmetry transformation \cite{21}.

\section{$D=4$ Supergravity from minimal Maxwell superalgebra $s\mathcal{M}%
_{4}$}

It was shown in Ref. \cite{15} that the minimal Maxwell superalgebra
$s\mathcal{M}_{4}$ in $D=4$ can be found by an $S$-expansion of the
$\mathfrak{osp}\left(  4|1\right)  $ superalgebra given by $\left(
\ref{ADS01}\right)  -\left(  \ref{ADS05}\right)  $. \ In fact, following
\cite{15} let us consider the $S$-expansion of the Lie superalgebra
$\mathfrak{osp}\left(  4|1\right)  $ using $S_{E}^{\left(  4\right)
}=\left\{  \lambda_{0},\lambda_{1},\lambda_{2},\lambda_{3},\lambda_{4}%
,\lambda_{5}\right\}  $ as an abelian semigroup. \ The elements of the
semigroup are dimensionless and obey the multiplication law%
\begin{equation}
\lambda_{\alpha}\lambda_{\beta}=\left\{
\begin{array}
[c]{c}%
\lambda_{\alpha+\beta}\text{, \ \ \ \ when }\alpha+\beta\leq\lambda_{5},\\
\lambda_{5}\text{, \ \ \ \ \ \ \ when }\alpha+\beta>\lambda_{5},
\end{array}
\right.
\end{equation}
where $\lambda_{5}$ plays the role of the zero element of the
semigroup.\ \ After extracting a resonant subalgebra and considering a
reduction, one finds a new algebra, whose generators $J_{ab}=\lambda_{0}%
\tilde{J}_{ab},$\ $P_{a}=\lambda_{2}\tilde{P}_{a},$\ $\tilde{Z}_{ab}%
=\lambda_{2}\tilde{J}_{ab},$ $Z_{ab}=\lambda_{4}\tilde{J}_{ab},$ $Q_{\alpha
}=\lambda_{1}\tilde{Q}_{\alpha},$ $\Sigma_{\alpha}=\lambda_{3}\tilde
{Q}_{\alpha}$ satisfy the following commutation relations%
\begin{align}
\left[  J_{ab},J_{cd}\right]   &  =\eta_{bc}J_{ad}-\eta_{ac}J_{bd}-\eta
_{bd}J_{ac}+\eta_{ad}J_{bc},\\
\left[  J_{ab},P_{c}\right]   &  =\eta_{bc}P_{a}-\eta_{ac}P_{b},\text{
\ \ \ \ \ \ \ }\left[  P_{a},P_{b}\right]  =Z_{ab},\\
\left[  J_{ab},Z_{cd}\right]   &  =\eta_{bc}Z_{ad}-\eta_{ac}Z_{bd}-\eta
_{bd}Z_{ac}+\eta_{ad}Z_{bc},\\
\left[  P_{a},Q_{\alpha}\right]   &  =-\frac{1}{2}\left(  \gamma_{a}%
\Sigma\right)  _{\alpha},\\
\left[  J_{ab},Q_{\alpha}\right]   &  =-\frac{1}{2}\left(  \gamma
_{ab}Q\right)  _{\alpha},\\
\left[  J_{ab},\Sigma_{\alpha}\right]   &  =-\frac{1}{2}\left(  \gamma
_{ab}\Sigma\right)  _{\alpha},\\
\left\{  Q_{\alpha},Q_{\beta}\right\}   &  =-\frac{1}{2}\left[  \left(
\gamma^{ab}C\right)  _{\alpha\beta}\tilde{Z}_{ab}-2\left(  \gamma^{a}C\right)
_{\alpha\beta}P_{a}\right]  ,\\
\left\{  Q_{\alpha},\Sigma_{\beta}\right\}   &  =-\frac{1}{2}\left(
\gamma^{ab}C\right)  _{\alpha\beta}Z_{ab},
\end{align}%
\begin{align}
\left[  J_{ab},\tilde{Z}_{ab}\right]   &  =\eta_{bc}\tilde{Z}_{ad}-\eta
_{ac}\tilde{Z}_{bd}-\eta_{bd}\tilde{Z}_{ac}+\eta_{ad}\tilde{Z}_{bc},\\
\left[  \tilde{Z}_{ab},\tilde{Z}_{cd}\right]   &  =\eta_{bc}Z_{ad}-\eta
_{ac}Z_{bd}-\eta_{bd}Z_{ac}+\eta_{ad}Z_{bc},\\
\left[  \tilde{Z}_{ab},Q_{\alpha}\right]   &  =-\frac{1}{2}\left(  \gamma
_{ab}\Sigma\right)  _{\alpha},\\
\text{others}  &  =0.
\end{align}
\ This new superalgebra obtained after a reduced resonant $S$-expansion of
$\mathfrak{osp}\left(  4|1\right)  $ superalgebra corresponds to a minimal
superMaxwell algebra $s\mathcal{M}_{4}$ in $D=4$ which contains the Maxwell
algebra $\mathcal{M}_{4}=\left\{  J_{ab},P_{a},Z_{ab}\right\}  $ and the
Lorentz type subalgebra $\mathcal{L}^{\mathcal{M}_{4}}=\left\{  J_{ab}%
,Z_{ab}\right\}  $ as subalgebras. \ One can see that setting $\tilde{Z}%
_{ab}=0$ leads us to the minimal Maxwell superalgebra introduced in Ref.
\cite{14}. \ This can be done since the Jacobi identities for spinors
generators are satisfied due to the gamma matrix identity $\left(  C\gamma
^{a}\right)  _{\left(  \alpha\beta\right.  }\left(  C\gamma_{a}\right)
_{\left.  \gamma\delta\right)  }=0$ $\left(  \text{cyclic permutations of
}\alpha,\beta,\gamma\right)  $. \ An alternative expansion method to obtain
the minimal Maxwell superalgebra can be found in Ref. \cite{7}.

\ In order to write down an action for $s\mathcal{M}_{4}$, we start from the
one-form gauge connection%
\begin{equation}
A=\frac{1}{2}\omega^{ab}J_{ab}+\frac{1}{2}\tilde{k}^{ab}\tilde{Z}_{ab}%
+\frac{1}{2}k^{ab}Z_{ab}+\frac{1}{l}e^{a}P_{a}+\frac{1}{\sqrt{l}}\psi^{\alpha
}Q_{\alpha}+\frac{1}{\sqrt{l}}\xi^{\alpha}\Sigma_{\alpha},
\end{equation}
where the 1-form gauge fields are given by%
\[%
\begin{tabular}
[c]{ll}%
$\omega^{ab}=\omega^{\left(  ab,0\right)  }=\lambda_{0}\tilde{\omega}^{ab},$ &
$\ e^{a}=e^{\left(  a,2\right)  }=\lambda_{2}\tilde{e}^{a},$\\
$\tilde{k}^{ab}=\omega^{\left(  ab,2\right)  }=\lambda_{2}\tilde{\omega}%
^{ab},$ & $\ \psi^{\alpha}=\psi^{\left(  \alpha,1\right)  }=\lambda_{1}%
\tilde{\psi}^{\alpha},$\\
$k^{ab}=\omega^{\left(  ab,4\right)  }=\lambda_{4}\tilde{\omega}^{ab},$ &
$\ \xi^{\alpha}=\psi^{\left(  \alpha,3\right)  }=\lambda_{3}\tilde{\psi
}^{\alpha},$%
\end{tabular}
\
\]
in terms of $\tilde{e}^{a},\tilde{\omega}^{ab}$ and $\tilde{\psi}$ which are
the components of the $\mathfrak{osp}\left(  4|1\right)  $ connection.

The associated two-form curvature $F=dA+A\wedge A$ is%
\begin{equation}
F=F^{A}T_{A}=\frac{1}{2}R^{ab}J_{ab}+\frac{1}{l}R^{a}P_{a}+\frac{1}{2}%
\tilde{F}^{ab}\tilde{Z}_{ab}+\frac{1}{2}F^{ab}Z_{ab}+\frac{1}{\sqrt{l}}%
\Psi^{\alpha}Q_{\alpha}+\frac{1}{\sqrt{l}}\Xi^{\alpha}\Sigma_{\alpha},
\label{curm}%
\end{equation}
where%
\begin{align}
R^{ab}  &  =d\omega^{ab}+\omega_{\text{ }c}^{a}\omega^{cb},\\
R^{a}  &  =de^{a}+\omega_{\text{ }b}^{a}e^{b}-\frac{1}{2}\bar{\psi}\gamma
^{a}\psi,\\
\tilde{F}^{ab}  &  =d\tilde{k}^{ab}+\omega_{\text{ }c}^{a}\tilde{k}%
^{cb}-\omega_{\text{ }c}^{b}\tilde{k}^{ca}+\frac{1}{2l}\bar{\psi}\gamma
^{ab}\psi,\\
F^{ab}  &  =dk^{ab}+\omega_{\text{ }c}^{a}k^{cb}-\omega_{\text{ }c}^{b}%
k^{ca}+\tilde{k}_{\text{ }c}^{a}\tilde{k}^{cb}+\frac{1}{l^{2}}e^{a}e^{b}%
+\frac{1}{l}\bar{\xi}\gamma^{ab}\psi,\\
\Psi &  =d\psi+\frac{1}{4}\omega_{ab}\gamma^{ab}\psi=D\psi,\\
\Xi &  =d\xi+\frac{1}{4}\omega_{ab}\gamma^{ab}\xi+\frac{1}{4}\tilde{k}%
_{ab}\gamma^{ab}\psi+\frac{1}{2l}e^{a}\gamma_{a}\psi\nonumber\\
&  =D\xi+\frac{1}{4}\tilde{k}_{ab}\gamma^{ab}\psi+\frac{1}{2l}e^{a}\gamma
_{a}\psi.
\end{align}
From the Bianchi identity $\nabla F=0$, where $\nabla=d+\left[  A,\cdot
\right]  $, it is possible to show that the Lorentz covariant exterior
derivatives of the curvatures are given by,
\begin{align}
DR^{ab}  &  =0,\\
DR^{a}  &  =R_{\text{ }b}^{a}e^{b}+\bar{\psi}\gamma^{a}\Psi,\\
D\tilde{F}^{ab}  &  =R_{\text{ }c}^{a}\tilde{k}^{cb}-R_{\text{ }c}^{b}%
\tilde{k}^{ca}-\frac{1}{l}\bar{\psi}\gamma^{ab}\Psi\\
DF^{ab}  &  =R_{\text{ }c}^{a}k^{cb}-R_{\text{ }c}^{b}k^{ca}+\tilde{F}_{\text{
}c}^{a}\tilde{k}^{cb}-\tilde{F}_{\text{ }c}^{b}\tilde{k}^{ca}+\frac{1}{l^{2}%
}R^{a}e^{b}-\frac{1}{l^{2}}e^{a}R^{b}\\
&  +\frac{1}{l}\bar{\Xi}\gamma^{ab}\psi-\frac{1}{l}\bar{\xi}\gamma^{ab}\Psi,\\
D\Psi &  =\frac{1}{4}R_{ab}\gamma^{ab}\psi,\\
D\Xi &  =\frac{1}{4}R_{ab}\gamma^{ab}\xi-\frac{1}{4}\tilde{k}_{ab}\gamma
^{ab}\Psi+\frac{1}{4}\tilde{F}_{ab}\gamma^{ab}\psi+\frac{1}{2l}R^{a}\gamma
_{a}\psi-\frac{1}{2l}e^{a}\gamma_{a}\Psi.
\end{align}
Then, the action can be written as
\begin{equation}
S=2\int\left\langle F\wedge F\right\rangle =2\int F^{A}\wedge F^{B}%
\left\langle T_{A}T_{B}\right\rangle , \label{actm}%
\end{equation}
where $\left\langle T_{A}T_{B}\right\rangle $ corresponds to an $S$-expanded
invariant tensor which is obtained from $\left(  \ref{Tinvads}\right)  $.
Using Theorem VII.1 of Ref.\cite{9} it is possible to show that these
components are given by
\begin{align}
\left\langle J_{ab}J_{cd}\right\rangle _{s\mathcal{M}_{4}}  &  =\alpha
_{0}\left\langle J_{ab}J_{cd}\right\rangle ,\label{Tinv01}\\
\left\langle J_{ab}\tilde{Z}_{cd}\right\rangle _{s\mathcal{M}_{4}}  &
=\alpha_{2}\left\langle J_{ab}J_{cd}\right\rangle ,\\
\left\langle \tilde{Z}_{ab}\tilde{Z}_{cd}\right\rangle _{s\mathcal{M}_{4}}  &
=\alpha_{4}\left\langle J_{ab}J_{cd}\right\rangle ,\label{INV}\\
\left\langle J_{ab}Z_{cd}\right\rangle _{s\mathcal{M}_{4}}  &  =\alpha
_{4}\left\langle J_{ab}J_{cd}\right\rangle ,\\
\left\langle Q_{\alpha}Q_{\beta}\right\rangle _{s\mathcal{M}_{4}}  &
=\alpha_{2}\left\langle Q_{\alpha}Q_{\beta}\right\rangle ,\\
\left\langle Q_{\alpha}\Sigma_{\beta}\right\rangle _{s\mathcal{M}_{4}}  &
=\alpha_{4}\left\langle Q_{\alpha}Q_{\beta}\right\rangle , \label{Tinv06}%
\end{align}
where%
\begin{align}
\left\langle J_{ab}J_{cd}\right\rangle  &  =\epsilon_{abcd}\\
\left\langle Q_{\alpha}Q_{\beta}\right\rangle  &  =2\left(  \gamma_{5}\right)
_{\alpha\beta}%
\end{align}
and the $\alpha$'s are dimensionless arbitrary independent constants.

Considering the different components of the invariant tensor $\left(
\ref{Tinv01}\right)  -\left(  \ref{Tinv06}\right)  $ and the two-form
curvature $\left(  \ref{curm}\right)  $, we found that the action can be
written as%
\begin{align}
S  &  =2\int\left(  \frac{1}{4}\alpha_{0}\epsilon_{abcd}R^{ab}R^{cd}+\frac
{1}{2}\alpha_{2}\epsilon_{abcd}R^{ab}\tilde{F}^{cd}+\frac{1}{2}\alpha
_{4}\epsilon_{abcd}R^{ab}F^{cd}\right. \nonumber\\
&  \left.  +\frac{1}{4}\alpha_{4}\epsilon_{abcd}\tilde{F}^{ab}\tilde{F}%
^{cd}+\frac{2}{l}\alpha_{2}\bar{\Psi}\gamma_{5}\Psi+\frac{4}{l}\alpha_{4}%
\bar{\Psi}\gamma_{5}\Xi\right)  \label{general}%
\end{align}
or explicitly,
\begin{align}
S  &  =\int\frac{\alpha_{0}}{2}\epsilon_{abcd}R^{ab}R^{cd}+\alpha_{2}%
\epsilon_{abcd}\left(  R^{ab}D\tilde{k}^{cd}+\frac{1}{2l}R^{ab}\bar{\psi
}\gamma^{cd}\psi\right) \nonumber\\
&  +\frac{4}{l}\alpha_{2}D\bar{\psi}\gamma_{5}D\psi+\alpha_{4}\epsilon
_{abcd}\left(  R^{ab}Dk^{cd}+\frac{1}{2}D\tilde{k}^{ab}D\tilde{k}^{cd}%
+\frac{1}{l^{2}}R^{ab}e^{c}e^{d}\right. \nonumber\\
&  \left.  +\frac{1}{2l}D\tilde{k}^{ab}\bar{\psi}\gamma^{cd}\psi+R^{ab}%
\tilde{k}_{\text{ }f}^{c}\tilde{k}^{fd}+\frac{1}{l}R^{ab}\bar{\xi}\gamma
^{cd}\psi\right) \nonumber\\
&  +\frac{8}{l}\alpha_{4}D\bar{\psi}\gamma_{5}D\xi+\frac{2}{l}\alpha_{4}%
D\bar{\psi}\gamma_{5}\tilde{k}_{ab}\gamma^{ab}\psi+\frac{4}{l^{2}}\alpha
_{4}\bar{\psi}e^{a}\gamma_{a}\gamma_{5}D\psi
\end{align}
where $D=d+\left[  \omega,\cdot\right]  .$ Using the gravitino Bianchi
identity%
\begin{equation}
D\Psi=\frac{1}{4}R^{ab}\gamma_{ab}\psi, \label{Bianchi}%
\end{equation}
and the gamma matrix identity%
\begin{equation}
2\gamma_{ab}\gamma_{5}=-\epsilon_{abcd}\gamma^{cd}, \label{gamma}%
\end{equation}
it is possible to show that,%
\begin{align*}
\frac{1}{2}\epsilon_{abcd}R^{ab}\bar{\psi}\gamma^{ab}\psi+4D\bar{\psi}%
\gamma_{5}D\psi &  =d\left(  4D\bar{\psi}\gamma_{5}\psi\right)  ,\\
\epsilon_{abcd}R^{ab}\bar{\xi}\gamma^{cd}\psi+8D\bar{\xi}\gamma_{5}D\psi &
=d\left(  8D\bar{\xi}\gamma_{5}\psi\right)  ,\\
\frac{1}{2}\epsilon_{abcd}D\tilde{k}^{ab}\bar{\psi}\gamma^{cd}\psi+2\bar{\psi
}\tilde{k}^{ab}\gamma_{ab}\gamma_{5}D\psi &  =d\left(  \bar{\psi}\tilde
{k}^{ab}\gamma_{ab}\gamma_{5}\psi\right)  .
\end{align*}
Thus the Mac Dowell-Mansouri like action for the $s\mathcal{M}_{4}$
superalgebra is finally given by%
\begin{align}
S  &  =\int\frac{\alpha_{0}}{2}\epsilon_{abcd}R^{ab}R^{cd}+\alpha_{2}d\left(
\epsilon_{abcd}R^{ab}\tilde{k}^{cd}+\frac{4}{l}D\bar{\psi}\gamma_{5}%
\psi\right) \nonumber\\
&  +\alpha_{4}\left[  \frac{1}{l^{2}}\epsilon_{abcd}R^{ab}e^{c}e^{d}+\frac
{4}{l^{2}}\bar{\psi}e^{a}\gamma_{a}\gamma_{5}D\psi\right. \nonumber\\
&  \left.  +d\left(  \epsilon_{abcd}\left(  R^{ab}k^{cd}+\frac{1}{2}D\tilde
{k}^{ab}\tilde{k}^{cd}\right)  +\frac{8}{l}\bar{\xi}\gamma_{5}D\psi+\frac
{1}{l}\bar{\psi}\tilde{k}^{ab}\gamma_{ab}\gamma_{5}\psi\right)  \right]
\label{sm4final}%
\end{align}
Here we can see that the lagrangian is split into three independent pieces
proportional to $\alpha_{0}$, $\alpha_{2}$ and $\alpha_{4}.$ \ The term
proportional to $\alpha_{0}$ corresponds to the Euler invariant. \ The piece
proportional to $\alpha_{2}$ is a boundary term. \ The term proportional to
$\alpha_{4}$ contains the Einstein-Hilbert term $\epsilon_{abcd}R^{ab}%
e^{c}e^{d}$ plus the Rarita-Schwinger lagrangian $4\bar{\psi}e^{a}\gamma
_{a}\gamma_{5}D\psi$, and a boundary term.

From $\left(  \ref{sm4final}\right)  $ we can see that the minimal Maxwell
superalgebra $s\mathcal{M}_{4}$ leads us to the pure supergravity action plus
boundary terms. \ In this way the new Maxwell gauge fields do not contribute
to the dynamics and enlarge only the boundary terms. \ Furthermore, as a
consequence of the $S$-expansion procedure the supersymmetric cosmological
term disappears completely from the action for $s\mathcal{M}_{4}$.

This result is particularly interesting since it corresponds to the
supersymmetric case of Refs. \cite{10, 12}, where Einstein-Hilbert action is
obtained from Maxwell algebra%
\footnote{also known as
$\mathfrak{B}_{4}$ algebra}
as a Born-Infeld like action.

Let us note that if we consider $\tilde{k}^{ab}=0$, the term proportional to
$\alpha_{4}$ corresponds to the action found in \cite{18}, namely%
\begin{equation}
S|_{\tilde{k}^{ab}=0}=\alpha_{4}\int\frac{1}{l^{2}}\left(  \epsilon
_{abcd}R^{ab}e^{c}e^{d}+4\bar{\psi}e^{a}\gamma_{a}\gamma_{5}D_{\omega}%
\psi\right)  +d\left(  \epsilon_{abcd}R^{ab}k^{cd}+\frac{8}{l}\bar{\xi}%
\gamma_{5}D_{\omega}\psi\right)  \label{minimal}%
\end{equation}
which corresponds to four-dimensional pure supergravity plus a boundary term.
\ This is not a surprise but something expected, because as we said before
setting $\tilde{Z}_{ab}=0$ in $s\mathcal{M}_{4}$ leads us to the simplest
minimal Maxwell superalgebra \cite{15}, whose two-form curvature associated
allows the construction of $\left(  \ref{minimal}\right)  $ as was shown in
\cite{18}.

It would be interesting to study the possibility to obtain the action $\left(
\ref{sm4final}\right)  $ from the approach considered in Ref. \cite{22} in
which $N=1$ and $N=2$ supergravities are constructed in the presence of a non
trivial boundary.\

\subsection{$s\mathcal{M}_{4}$ gauge transformations and supersymmetry}

\qquad The gauge transformation of the connection $A$ is
\begin{equation}
\delta_{\rho}A=D\rho=d\rho+\left[  A,\rho\right]
\end{equation}
where $\rho$ is the $s\mathcal{M}_{4}$ gauge parameter,%
\begin{equation}
\rho=\frac{1}{2}\rho^{ab}J_{ab}+\frac{1}{2}\tilde{\kappa}^{ab}\tilde{Z}%
_{ab}+\frac{1}{2}\kappa^{ab}Z_{ab}+\frac{1}{l}\rho^{a}P_{a}+\frac{1}{\sqrt{l}%
}\epsilon^{\alpha}Q_{\alpha}+\frac{1}{\sqrt{l}}\varrho^{\alpha}\Sigma_{\alpha
}.
\end{equation}
Then, using%

\begin{equation}
\delta\left(  A^{A}T_{A}\right)  =d\rho+\left[  A^{B}T_{B},\rho^{C}%
T_{C}\right]  ,
\end{equation}
the $s\mathcal{M}_{4}$ gauge transformation are given by%
\begin{align}
\delta\omega^{ab}  &  =D\rho^{ab},\label{ST01}\\
\delta\tilde{k}^{ab}  &  =D\tilde{\kappa}^{ab}-\left(  \tilde{k}_{\text{
}c\text{ }}^{a}\rho_{\text{ }c}^{b}-\tilde{k}^{bc}\rho_{\text{ }c}^{a}\right)
-\frac{1}{l}\bar{\epsilon}\gamma^{ab}\psi,\\
\delta k^{ab}  &  =D\kappa^{ab}-\left(  k^{ac}\rho_{\text{ }c}^{b}-k^{bc}%
\rho_{\text{ }c}^{a}\right)  -\left(  \tilde{k}^{ac}\tilde{\kappa}_{\text{ }%
c}^{b}-\tilde{k}^{bc}\tilde{\kappa}_{\text{ }c}^{a}\right) \nonumber\\
&  +\frac{2}{l^{2}}e^{a}\rho^{b}-\frac{1}{l}\bar{\varrho}\gamma^{ab}\psi
-\frac{1}{l}\bar{\epsilon}\gamma^{ab}\xi,\\
\delta e^{a}  &  =D\rho^{a}\,+e^{b}\rho_{b}^{\text{ }a}+\bar{\epsilon}%
\gamma^{a}\psi,\\
\delta\psi &  =d\epsilon+\frac{1}{4}\omega^{ab}\gamma_{ab}\epsilon-\frac{1}%
{4}\rho^{ab}\gamma_{ab}\psi,\\
\delta\xi &  =d\varrho+\frac{1}{4}\omega^{ab}\gamma_{ab}\varrho+\frac{1}%
{2l}e^{a}\gamma_{a}\epsilon-\frac{1}{2l}\rho^{a}\gamma_{a}\psi-\frac{1}{4}%
\rho^{ab}\gamma_{ab}\xi\nonumber\label{ST06}\\
&  +\frac{1}{4}\tilde{k}^{ab}\gamma_{ab}\epsilon-\frac{1}{4}\tilde{\kappa
}^{ab}\gamma_{ab}\psi.
\end{align}
In the same way, from the gauge variation of the curvature%
\begin{equation}
\delta_{\rho}F=\left[  F,\rho\right]
\end{equation}
it is possible to show that the gauge transformations of the curvature $F$ are
given by%
\begin{align}
\delta R^{ab}  &  =R^{ac}\rho_{c}^{\text{ }b}-R^{cb}\rho_{\text{ }c}^{a},\\
\delta\tilde{F}^{ab}  &  =\left(  R^{ac}\tilde{\kappa}_{c}^{\text{ }b}%
-R^{bc}\tilde{\kappa}_{\text{ }c}^{a}\right)  -\left(  \tilde{F}^{ac}%
\rho_{\text{ }c}^{b}-\tilde{F}^{bc}\rho_{\text{ }c}^{a}\right)  -\frac{1}%
{l}\bar{\epsilon}\gamma^{ab}\Psi,\\
\delta F^{ab}  &  =\left(  R^{ac}\kappa_{c}^{\text{ }b}-R^{bc}\kappa_{\text{
}c}^{a}\right)  -\left(  F^{ac}\rho_{\text{ }c}^{b}-F^{bc}\rho_{\text{ }c}%
^{a}\right)  -\left(  \tilde{F}^{ac}\tilde{\kappa}_{\text{ }c}^{b}-\tilde
{F}^{ac}\tilde{\kappa}_{\text{ }c}^{a}\right) \nonumber\\
&  +\frac{2}{l^{2}}R^{a}\rho^{b}-\frac{1}{l}\bar{\varrho}\gamma^{ab}\Psi
-\frac{1}{l}\bar{\epsilon}\gamma^{ab}\Xi,\\
\delta R^{a}  &  =R_{\text{ }b}^{a}\rho^{b}+R^{b}\rho_{b}^{\text{ }a}%
+\bar{\epsilon}\gamma^{a}\Psi,\\
\delta\Psi &  =\frac{1}{4}R^{ab}\gamma_{ab}\epsilon-\frac{1}{4}\rho^{ab}%
\gamma_{ab}\Psi,\\
\delta\Xi &  =\frac{1}{4}R^{ab}\gamma_{ab}\varrho+\frac{1}{2l}R^{a}\gamma
_{a}\epsilon-\frac{1}{2l}\rho^{a}\gamma_{a}\Psi-\frac{1}{4}\rho^{ab}%
\gamma_{ab}\Xi+\frac{1}{4}\tilde{F}^{ab}\gamma_{ab}\epsilon-\frac{1}{4}%
\tilde{\kappa}^{ab}\gamma_{ab}\Psi,
\end{align}

Although the Mac Dowell-Mansouri like action $\left(  \ref{sm4final}\right)  $
is built from the $s\mathcal{M}_{4}$ curvature, it is \textbf{not} invariant
under the $s\mathcal{M}_{4}$ gauge transformations. \ As we can see the action
does not correspond to a Yang-Mills action, nor a topological invariant.

Furthermore, the action is not invariant under gauge supersymmetry. \ In fact,
if we consider the variation of the action $\left(  \ref{sm4final}\right)  $
under gauge supersymmetry , we find
\begin{equation}
\delta_{susy}S=-\frac{4}{l^{2}}\alpha_{4}\int R^{a}\bar{\Psi}\gamma_{a}%
\gamma_{5}\epsilon.
\end{equation}

As in $\mathfrak{osp}\left(  4|1\right)  $ and super-Poincar\'{e} cases, the
action is invariant under gauge supersymmetry imposing the super torsion
constraint%
\begin{equation}
R^{a}=0.
\end{equation}
This yields to express the spin connection $\omega^{ab}$ in terms of the
vielbein and the gravitino fields. \ This leads to the supersymmetric action
for $s\mathcal{M}_{4}$ superalgebra in second order formalism.

Alternatively, it is possible to have supersymmetry in first order formalism
if we modify the supersymmetry transformation for the spin connection
$\omega^{ab}$. \ In fact, if we consider the variation of the action under an
arbitrary $\delta\omega^{ab}$ we find%
\begin{equation}
\delta_{\omega}S=\frac{2}{l^{2}}\alpha_{4}\int\epsilon_{abcd}R^{a}e^{b}%
\delta\omega^{cd},
\end{equation}
thus the variation vanish for arbitrary $\delta\omega^{ab}$ if $R^{a}=0$.
\ Similarly to Ref. \cite{21}, it is possible to modify $\delta\omega^{ab}$
adding an extra piece to the gauge transformation such that the variation of
the action can be written as%
\begin{equation}
\delta S=-\frac{4}{l^{2}}\alpha_{4}\int R^{a}\left(  \bar{\Psi}\gamma
_{a}\gamma_{5}\epsilon-\frac{1}{2}\epsilon_{abcd}e^{b}\delta_{extra}%
\omega^{cd}\right)  .
\end{equation}
In order to have an invariant action, $\delta_{extra}\omega^{ab}$ is given by%
\begin{equation}
\delta_{extra}\omega^{ab}=2\epsilon^{abcd}\left(  \bar{\Psi}_{ec}\gamma
_{d}\gamma_{5}\epsilon+\bar{\Psi}_{de}\gamma_{c}\gamma_{5}\epsilon-\bar{\Psi
}_{cd}\gamma_{e}\gamma_{5}\epsilon\right)  e^{e},
\end{equation}
with $\bar{\Psi}=\bar{\Psi}_{ab}e^{a}e^{b}$.

Then the action in the first order formalism is invariant under the following
supersymmetry transformations%
\begin{align}
\delta\omega^{ab}  &  =2\epsilon^{abcd}\left(  \bar{\Psi}_{ec}\gamma_{d}%
\gamma_{5}\epsilon+\bar{\Psi}_{de}\gamma_{c}\gamma_{5}\epsilon-\bar{\Psi}%
_{cd}\gamma_{e}\gamma_{5}\epsilon\right)  e^{e},\\
\delta\tilde{k}^{ab}  &  =-\frac{1}{l}\bar{\epsilon}\gamma^{ab}\psi,\\
\delta k^{ab}  &  =-\frac{1}{l}\bar{\epsilon}\gamma^{ab}\xi,\\
\delta e^{a}  &  =\bar{\epsilon}\gamma^{a}\psi,\\
\delta\psi &  =d\epsilon+\frac{1}{4}\omega^{ab}\gamma_{ab}\epsilon
=D\epsilon,\\
\delta\xi &  =\frac{1}{2l}e^{a}\gamma_{a}\epsilon+\frac{1}{4}\tilde{k}%
^{ab}\gamma_{ab}\epsilon.
\end{align}
On the other hand, it is important to note that there is a new supersymmetry
related to the spinor charge $\Sigma.$ The new supersymmetry transformations
are given by%
\begin{align}
\delta\omega^{ab}  &  =0,\\
\delta\tilde{k}^{ab}  &  =0,\\
\delta k^{ab}  &  =-\frac{1}{l}\bar{\varrho}\gamma^{ab}\psi,\\
\delta e^{a}  &  =0,\\
\delta\psi &  =0,\\
\delta\xi &  =d\varrho+\frac{1}{4}\omega^{ab}\gamma_{ab}\varrho.
\end{align}
Considering the variation of the action $\left(  \ref{sm4final}\right)  $
under the new gauge supersymmetry transformations, we find that the action is
truly invariant%
\begin{equation}
\delta S=0.
\end{equation}
Then one can see that the action is off-shell invariant under a subalgebra of
$s\mathcal{M}_{4}$ given by $s\mathcal{L}_{\mathcal{M}_{4}}=\left\{
J_{ab},\tilde{Z}_{ab},Z_{ab},\Sigma_{\alpha}\right\}  $ which corresponds to a
Lorentz type superalgebra. \ These results are interesting since we have shown
that the Poincar\'{e} supersymmetry is not the only supersymmetry of $D=4$
pure supergravity.

\section{$D=4$ Supergravity from the minimal Maxwell superalgebra type
$s\mathcal{M}_{m+2}$}

\ \qquad It was shown in Ref. \cite{15} that the $D=4$ minimal superMaxwell
algebra type $s\mathcal{M}_{m+2}$ can be found by an $S$-expansion of the
$\mathfrak{osp}\left(  4|1\right)  $ superalgebra given by $\left(
\ref{ADS01}\right)  -\left(  \ref{ADS05}\right)  $. \ In fact, following
\cite{15} let us consider the $S$-expansion of the Lie superalgebra
$\mathfrak{osp}\left(  4|1\right)  $ using $S_{E}^{\left(  2m\right)
}=\left\{  \lambda_{0},\lambda_{1},\lambda_{2},\cdots,\lambda_{2m+1}\right\}
$ as the relevant finite abelian semigroup. \ The elements of the semigroup
are dimensionless and obey the multiplication law%
\begin{equation}
\lambda_{\alpha}\lambda_{\beta}=\left\{
\begin{array}
[c]{c}%
\lambda_{\alpha+\beta}\text{, \ \ \ \ \ \ \ \ when }\alpha+\beta\leq
\lambda_{2m+1},\\
\lambda_{2m+1}\text{, \ \ \ \ \ \ \ when }\alpha+\beta>\lambda_{2m+1},
\end{array}
\right.  \label{lm03}%
\end{equation}
where $\lambda_{2m+1}$ plays the role of the zero element of the semigroup
$S_{E}^{\left(  2m\right)  }$. \ After extracting a resonant subalgebra and
considering a reduction, one finds the $D=4$ minimal Maxwell superalgebra type
$s\mathcal{M}_{m+2}$. \ The new superalgebra obtained by the $S$-expansion
procedure is generated by%
\begin{equation}
\left\{  J_{ab,\left(  k\right)  },P_{a,\left(  l\right)  },Q_{\alpha,\left(
k\right)  }\right\}  ,
\end{equation}
where these new generators can be written as%
\begin{align}
J_{ab,\left(  k\right)  }  &  =\lambda_{2k}\tilde{J}_{ab},\label{Gsm1}\\
P_{a,\left(  l\right)  }  &  =\lambda_{2l}\tilde{P}_{a},\text{\ }\\
Q_{\alpha,\left(  p\right)  }  &  =\lambda_{2p-1}\tilde{Q}_{\alpha},
\label{Gsm3}%
\end{align}
with $k=0,\dots,m$; $l=p=1,\dots,m$. Here, the generators $\tilde{J}%
_{ab},\tilde{P}_{a}$ and $\tilde{Q}_{\alpha}$ correspond to the
$\mathfrak{osp}\left(  4|1\right)  $ generators. \ The new generators satisfy
the commutation relations%
\begin{align}
\left[  J_{ab,\left(  k\right)  },J_{cd,\left(  j\right)  }\right]   &
=\eta_{bc}J_{ad,\left(  k+j\right)  }-\eta_{ac}J_{bd,\left(  k+j\right)
}-\eta_{bd}J_{ac,\left(  k+j\right)  }+\eta_{ad}J_{bc,\left(  k+j\right)  },\\
\left[  J_{ab,\left(  k\right)  },P_{a,\left(  l\right)  }\right]   &
=\eta_{bc}P_{a,\left(  k+l\right)  }-\eta_{ac}P_{b,\left(  k+l\right)  },\\
\left[  P_{a,\left(  l\right)  },P_{b,\left(  n\right)  }\right]   &
=J_{ab,\left(  l+n\right)  },\\
\left[  J_{ab,\left(  k\right)  },Q_{\alpha,\left(  p\right)  }\right]   &
=-\frac{1}{2}\left(  \gamma_{ab}Q\right)  _{\alpha,\left(  k+p\right)  },\\
\left[  P_{a,\left(  l\right)  },Q_{\alpha,\left(  p\right)  }\right]   &
=-\frac{1}{2}\left(  \gamma_{a}Q\right)  _{\alpha,\left(  l+p\right)  },\\
\left\{  Q_{\alpha,\left(  p\right)  },Q_{\beta,\left(  q\right)  }\right\}
&  =-\frac{1}{2}\left[  \left(  \gamma^{ab}C\right)  _{\alpha\beta
}J_{ab,\left(  p+q\right)  }-2\left(  \gamma^{a}C\right)  _{\alpha\beta
}P_{a,\left(  p+q\right)  }\right]  .
\end{align}
Naturally, when $k+j>m$ the generators $T_{A}^{\left(  k\right)  }$ and
$T_{B}^{\left(  j\right)  }$ are abelian. \ One sees that if we redefine the
generators as%
\[%
\begin{tabular}
[c]{ll}%
$J_{ab}=J_{ab,0}=\lambda_{0}\tilde{J}_{ab},$ & $P_{a}=P_{a,2}=\lambda
_{2}\tilde{P}_{a},$\\
$Z_{ab}^{\left(  k\right)  }=J_{ab,4k}=\lambda_{4k}\tilde{J}_{ab},$ &
$Z_{a}^{\left(  l\right)  }=P_{a,4l+2}=\lambda_{4l+2}\tilde{P}_{a},$\\
$\tilde{Z}_{ab}^{\left(  k\right)  }=J_{ab,4k-2}=\lambda_{4k-2}\tilde{J}%
_{ab},$ & $\tilde{Z}_{a}^{\left(  l\right)  }=P_{a,4l}=\lambda_{4l}\tilde
{P}_{a},$\\
$Q_{\alpha}=Q_{\alpha,1}=\lambda_{1}\tilde{Q}_{\alpha},$ & $\Sigma_{\alpha
}^{\left(  k\right)  }=Q_{\alpha,4k-1}=\lambda_{4k-1}\tilde{Q}_{\alpha},$\\
$\Phi_{\alpha}^{\left(  l\right)  }=Q_{\alpha,4k+1}=\lambda_{4k+1}\tilde
{Q}_{\alpha},$ & $\ \ \ \ \ $%
\end{tabular}
\ \ \
\]
we obtain the commutation relations for $s\mathcal{M}_{m+2}$ introduced in
\cite{15}. \ However, if we want to build an action and avoid extensive terms
we shall use $\left(  \ref{Gsm1}\right)  -\left(  \ref{Gsm3}\right)  $. \ In
order to write down a Lagrangian for $s\mathcal{M}_{m+2}$, we start from the
one-form gauge connection%
\begin{equation}
A=\frac{1}{2}\sum_{k}\omega^{ab,\left(  k\right)  }J_{ab,\left(  k\right)
}+\frac{1}{l}\sum_{l}e^{a,\left(  l\right)  }P_{a,\left(  l\right)  }+\frac
{1}{\sqrt{l}}\sum_{p}\psi^{\alpha,\left(  p\right)  }Q_{\alpha,\left(
p\right)  },
\end{equation}
where the different components are given by
\begin{align}
\omega^{ab,\left(  k\right)  }  &  =\lambda_{2k}\tilde{\omega}^{ab},\\
e^{a,\left(  l\right)  }  &  =\lambda_{2l}\tilde{e}^{a},\\
\psi^{\alpha,\left(  p\right)  }  &  =\lambda_{2p-1}\tilde{\psi}^{\alpha},
\end{align}
in terms of $\tilde{e}^{a},\tilde{\omega}^{ab}$ and $\tilde{\psi}$ which are
the components of the $\mathfrak{osp}\left(  4|1\right)  $ connection.

The associated two-form curvature $F=dA+A\wedge A$ is%
\begin{equation}
F=F^{A}T_{A}=\frac{1}{2}\sum_{k}\mathcal{R}^{ab,\left(  k\right)
}J_{ab,\left(  k\right)  }+\frac{1}{l}\sum_{l}R^{a,\left(  l\right)
}P_{a,\left(  l\right)  }+\frac{1}{\sqrt{l}}\sum_{p}\Psi^{\alpha,\left(
p\right)  }Q_{\alpha,\left(  p\right)  }, \label{2formSM}%
\end{equation}
where%
\begin{align}
\mathcal{R}^{ab,\left(  k\right)  }  &  =d\omega^{ab,\left(  k\right)
}+\omega_{\text{ }c}^{a\text{ }\left(  i\right)  }\wedge\omega^{cb,\left(
j\right)  }\delta_{i+j}^{k}+\frac{1}{l^{2}}e^{a,\left(  l\right)
}e^{b,\left(  n\right)  }\delta_{l+n}^{k}\nonumber\\
&  +\frac{1}{2l}\bar{\psi}^{\left(  p\right)  }\gamma^{ab}\wedge\psi^{\left(
q\right)  }\delta_{p+q}^{2k},\label{curvgen01}\\
R^{a,\left(  l\right)  }  &  =de^{a,\left(  l\right)  }+\omega_{\text{ }%
b}^{a\text{ }\left(  k\right)  }\wedge e^{b,\left(  n\right)  }\delta
_{k+n}^{l}-\frac{1}{2}\bar{\psi}^{\left(  p\right)  }\gamma^{a}\wedge
\psi^{\left(  q\right)  }\delta_{p+q}^{2l},\label{supertorsionexp}\\
\Psi^{\left(  p\right)  }  &  =d\psi^{\left(  p\right)  }+\frac{1}{4}%
\omega_{ab}^{\text{ \ \ }\left(  k\right)  }\gamma^{ab}\wedge\psi^{\left(
q\right)  }\delta_{k+q}^{p}+\frac{1}{2l}e^{a,\left(  l\right)  }\gamma
_{a}\wedge\psi^{\left(  q\right)  }\delta_{l+q}^{p}, \label{curvgen3}%
\end{align}
with $k=0,\dots,m$; $l=p=1,\dots,m$. \ The Bianchi identities can be obtained
by considering $\nabla F=0$, where $\nabla=d+\left[  A,\cdot\right]  $,%
\begin{align}
D\mathcal{R}^{ab,\left(  k\right)  }  &  =\left(  \mathcal{R}^{ac,\left(
i\right)  }\omega_{c}^{\text{ }b,\left(  j+1\right)  }-\mathcal{R}^{bc,\left(
i\right)  }\omega_{c}^{\text{ }a,\left(  j+1\right)  }\right)  \delta
_{i+j+1}^{k}\nonumber\\
&  +\frac{1}{l}\left(  R^{a,\left(  l\right)  }e^{b,\left(  n\right)
}-e^{a,\left(  n\right)  }R^{b,\left(  l\right)  }\right)  \delta_{l+n}%
^{k}-\frac{1}{l}\bar{\psi}^{\left(  p\right)  }\gamma^{ab}\Psi^{\left(
q\right)  }\delta_{p+q}^{2k},\\
DR^{a,\left(  l\right)  }  &  =\mathcal{R}^{ab,\left(  i\right)  }%
e_{b}^{\text{ \ },\left(  j\right)  }\delta_{i+j}^{l}+R^{c,\left(  n\right)
}\omega_{c}^{\text{ }a,\left(  j+1\right)  }\delta_{n+j+1}^{l}+\bar{\psi
}^{\left(  p\right)  }\gamma^{a}\Psi^{\left(  q\right)  }\delta_{p+q}^{2l},\\
D\Psi^{\left(  p\right)  }  &  =\frac{1}{4}\left(  \mathcal{R}^{ab,\left(
i\right)  }\gamma_{ab}\psi^{\left(  q\right)  }\right)  \delta_{i+q}^{p}%
-\frac{1}{4}\left(  \omega^{ab,\left(  i+1\right)  }\gamma_{ab}\Psi^{\left(
q\right)  }\right)  \delta_{i+1+q}^{p}\nonumber\\
&  +\frac{1}{2l}\left(  T^{a,\left(  l\right)  }\gamma_{a}\psi^{\left(
q\right)  }\right)  \delta_{l+q}^{p}-\frac{1}{2l}\left(  e^{a,\left(
l\right)  }\gamma_{a}\Psi^{\left(  q\right)  }\right)  \delta_{l+q}^{p},
\end{align}
where $D$ corresponds to the Lorentz covariant exterior derivative
$D=d+\left[  \omega,\cdot\right]  $.

Then the action can be written as
\begin{equation}
S=2\int\left\langle F\wedge F\right\rangle =2\int F^{A}\wedge F^{B}%
\left\langle T_{A}T_{B}\right\rangle ,
\end{equation}
where $\left\langle T_{A}T_{B}\right\rangle $ corresponds to an $S$-expanded
invariant tensor which is obtained from $\left(  \ref{Tinvads}\right)  $.
Using Theorem VII.1 of Ref. \cite{9} it is possible to show that these
components are given by%
\begin{align}
\left\langle J_{ab,\left(  k\right)  }J_{cd,\left(  j\right)  }\right\rangle
_{s\mathcal{M}_{m+2}}  &  =\alpha_{2\left(  k+j\right)  }\left\langle
J_{ab}J_{cd}\right\rangle ,\\
\left\langle Q_{\alpha,\left(  p\right)  }Q_{\beta,\left(  q\right)
}\right\rangle _{s\mathcal{M}_{m+2}}  &  =\alpha_{2\left(  p+q-1\right)
}\left\langle Q_{\alpha}Q_{\beta}\right\rangle ,
\end{align}
which can be written as%
\begin{align}
\left\langle J_{ab,\left(  k\right)  }J_{cd,\left(  j\right)  }\right\rangle
_{s\mathcal{M}_{m+2}}  &  =\alpha_{2\left(  k+j\right)  }\epsilon
_{abcd},\label{invSM01}\\
\left\langle Q_{\alpha,\left(  p\right)  }Q_{\beta,\left(  q\right)
}\right\rangle _{s\mathcal{M}_{m+2}}  &  =2\alpha_{2\left(  p+q-1\right)
}\left(  \gamma_{5}\right)  _{\alpha\beta}, \label{invSM02}%
\end{align}
where the $\alpha$'s are arbitrary independent constants and $J_{ab,\left(
k\right)  }$, $Q_{\alpha,\left(  p\right)  }$ are given by $\left(
\ref{Gsm1}\right)  ,\left(  \ref{Gsm3}\right)  $, respectively. \ Using the
different components of the invariant tensor $\left(  \ref{invSM01}\right)
-\left(  \ref{invSM02}\right)  $ and the two-form curvature $\left(
\ref{2formSM}\right)  $, we found that the action is given by%
\begin{equation}
S=2\int\sum_{k,j}\frac{\alpha_{2\left(  k+j\right)  }}{2}\epsilon
_{abcd}\mathcal{R}^{ab,\left(  k\right)  }\mathcal{R}^{cd,\left(  j\right)
}+\sum_{p,q}\alpha_{2\left(  p+q-1\right)  }\frac{4}{l}\bar{\Psi}^{\left(
p\right)  }\wedge\gamma_{5}\Psi^{\left(  q\right)  }, \label{sugrasMfinal}%
\end{equation}
with $k,j=0,\dots,m$; $p,q=1,\dots,m$.

\subsection{$s\mathcal{M}_{m+2}$ gauge transformations and supersymmetry}

\qquad The gauge transformation of the connection $A$ is
\begin{equation}
\delta_{\rho}A=D\rho=d\rho+\left[  A,\rho\right]
\end{equation}
where $\rho$ is the $s\mathcal{M}_{m+2}$ gauge parameter:%
\begin{equation}
\rho=\frac{1}{2}\sum_{k}\rho^{ab,\left(  k\right)  }J_{ab,\left(  k\right)
}+\frac{1}{l}\sum_{l}\rho^{a,\left(  l\right)  }P_{a,\left(  l\right)  }%
+\frac{1}{\sqrt{l}}\sum_{p}\epsilon^{\alpha,\left(  p\right)  }Q_{\alpha
,\left(  p\right)  }.
\end{equation}
Here we have written the components of the gauge parameter as an $S$-expansion
of the component of the $\mathfrak{osp}\left(  4|1\right)  $ gauge parameter,
\begin{align*}
\rho^{ab,\left(  k\right)  }  &  =\lambda_{2k}\tilde{\rho}^{ab},\\
\rho^{a,\left(  l\right)  }  &  =\lambda_{2l}\tilde{\rho}^{a},\\
\epsilon^{\alpha,\left(  p\right)  }  &  =\lambda_{2p-1}\tilde{\epsilon
}^{\alpha},
\end{align*}
with $k=0,\dots,m$; $l=p=1,\dots,m$ and $\lambda_{i}\in S_{E}^{\left(
2m\right)  }=\left\{  \lambda_{0},\lambda_{1},\lambda_{2},\cdots
,\lambda_{2m+1}\right\}  $. \ Then, using the multiplication law of the
semigroup $\left(  \ref{lm03}\right)  $ and%

\begin{equation}
\delta\left(  A^{A}T_{A}\right)  =d\rho+\left[  A^{B}T_{B},\rho^{C}%
T_{C}\right]  ,
\end{equation}
it is possible to show that the gauge transformations are given by%
\begin{align}
\delta\omega^{ab,\left(  k\right)  }  &  =D\rho^{ab,\left(  k\right)
}-\left(  \omega^{ac,\left(  i+1\right)  }\rho_{\text{ }c}^{b\text{ },\left(
j\right)  }-\omega^{bc,\left(  i+1\right)  }\rho_{\text{ }c}^{a\text{
},\left(  j\right)  }\right)  \delta_{i+j+1}^{k}\nonumber\\
&  +\frac{2}{l^{2}}e^{a,\left(  l\right)  }\rho^{b,\left(  n\right)  }%
\delta_{l+n}^{k}-\frac{1}{l}\bar{\epsilon}^{\left(  p\right)  }\gamma^{ab}%
\psi^{\left(  q\right)  }\delta_{p+q}^{2k},\\
\delta e^{a,\left(  l\right)  }  &  =D\rho^{a,\left(  l\right)  }%
+\omega_{\text{ }b}^{a\text{ },\left(  k+1\right)  }\rho^{b,\left(  n\right)
}\delta_{k+n+1}^{l}+e^{b,\left(  n\right)  }\rho_{b}^{\text{ }a,\left(
k\right)  }\delta_{n+k}^{l}+\bar{\epsilon}^{\left(  p\right)  }\gamma^{a}%
\psi^{\left(  q\right)  }\delta_{p+q}^{2l},\\
\delta\psi^{\left(  p\right)  }  &  =d\epsilon^{\left(  p\right)  }+\frac
{1}{4}\omega^{ab,\left(  k\right)  }\gamma_{ab}\epsilon^{\left(  q\right)
}\delta_{k+q}^{p}+\frac{1}{2l}e^{a,\left(  l\right)  }\gamma_{a}%
\epsilon^{\left(  q\right)  }\delta_{l+q}^{p}\nonumber\\
&  -\frac{1}{4}\rho^{ab,\left(  k\right)  }\gamma_{ab}\psi^{\left(  q\right)
}\delta_{k+q}^{p}-\frac{1}{2l}\rho^{a,\left(  l\right)  }\gamma_{a}%
\psi^{\left(  q\right)  }\delta_{l+q}^{p}.
\end{align}
In the same way, from the gauge variation of the curvature%
\begin{equation}
\delta_{\lambda}F=\left[  F,\lambda\right]
\end{equation}
it is possible to show that the gauge transformations of the curvature $F$ are
given by%
\begin{align}
\delta\mathcal{R}^{ab,\left(  k\right)  }  &  =\left(  \mathcal{R}^{ac,\left(
i\right)  }\rho_{c}^{\text{ }b,\left(  j\right)  }-\mathcal{R}^{cb,\left(
i\right)  }\rho_{\text{ }c}^{a\text{ },\left(  j\right)  }\right)
\delta_{i+j}^{k}+\frac{2}{l^{2}}R^{a,\left(  l\right)  }\rho^{b,\left(
n\right)  }\delta_{l+n}^{k}\nonumber\\
&  -\frac{1}{l}\bar{\epsilon}^{\left(  p\right)  }\gamma^{ab}\Psi^{\left(
q\right)  }\delta_{p+q}^{2k},\\
\delta R^{a,\left(  l\right)  }  &  =\mathcal{R}_{\text{ }b}^{a\text{
},\left(  k\right)  }\rho^{b,\left(  n\right)  }\delta_{k+n}^{l}+R^{b,\left(
n\right)  }\rho_{b}^{\text{ }a,\left(  k\right)  }\delta_{k+n}^{l}%
+\bar{\epsilon}^{\left(  p\right)  }\gamma^{a}\Psi^{\left(  q\right)  }%
\delta_{p+q}^{2l},\\
\delta\Psi^{\left(  p\right)  }  &  =\frac{1}{4}\mathcal{R}^{ab,\left(
k\right)  }\gamma_{ab}\epsilon^{\left(  q\right)  }\delta_{k+q}^{p}+\frac
{1}{2l}R^{a,\left(  l\right)  }\gamma_{a}\epsilon^{\left(  q\right)  }%
\delta_{l+q}^{p}-\frac{1}{4}\rho^{ab,\left(  k\right)  }\gamma_{ab}%
\Psi^{\left(  q\right)  }\delta_{k+q}^{p}\nonumber\\
&  -\frac{1}{2l}\rho^{a,\left(  l\right)  }\gamma_{a}\Psi^{\left(  q\right)
}\delta_{l+q}^{p},
\end{align}
with $k=i=j=0,\dots,m$; $l=n=p=q=1,\dots,m$.

Although the Mac Dowell-Mansouri like action $\left(  \ref{sugrasMfinal}%
\right)  $ is built from the $s\mathcal{M}_{m+2}$ curvature, it is
\textbf{not} invariant under the $s\mathcal{M}_{m+2}$ gauge transformations. \

Besides the action is not invariant under gauge supersymmetry. \ In fact, if
we consider the variation of the action $\left(  \ref{sugrasMfinal}\right)  $
under gauge supersymmetry related to $Q_{\left(  1\right)  }$, we find%
\begin{equation}
\delta_{susy}S=-\frac{4}{l^{2}}\int\sum_{k}\alpha_{2k}R^{a,\left(  l\right)
}\bar{\Psi}^{\left(  p\right)  }\gamma_{a}\gamma_{5}\epsilon\delta_{l+p}^{k},
\end{equation}
with $k=2,\dots,m$; $l,p\geq1$ and where $\epsilon$ is the gauge parameter
associated to the spinor charge $Q_{\left(  1\right)  }$.

As in the previous case the action is invariant for every value of $k$ under
gauge supersymmetry imposing the expanded super torsion constraint%
\begin{equation}
R^{a,\left(  l\right)  }=0.
\end{equation}
This yields to express the expanded spin connection $\omega^{ab,\left(
k\right)  }$ in terms of the expanded fields as we can see in $\left(
\ref{supertorsionexp}\right)  $. \ This leads to the supersymmetric action for
the $s\mathcal{M}_{m+2}$ superalgebra in the second order formalism.

Alternatively, since the $\alpha$'s are arbitrary and independent we can study
the supersymmetry in each term separately. Then if we consider the variation
of the action proportional to $\alpha_{2k}$ under gauge supersymmetry
transformations asociated to $Q_{\left(  k-1\right)  }$, we find%
\[
\delta_{susy}S=-\frac{4}{l^{2}}\alpha_{2k}\int R^{a}\bar{\Psi}\gamma_{a}%
\gamma_{5}\epsilon^{\left(  k-1\right)  },
\]
with $k=2,\dots,m$ and where $\epsilon^{\left(  k-1\right)  }$ is the gauge
parameter associated to the spinor charge $Q_{\left(  k-1\right)  }$. \ Here
$R^{a}$ and $\Psi$ correspond to $R^{a,\left(  1\right)  }$ and $\Psi^{\left(
1\right)  }$ respectively. \

It is possible to have invariance under supersymmetry in first order formalism
in every term if we modify the supersymmetry transformation for every expanded
spin connection. \ In fact, if we consider the variation of the action under
an arbitrary $\delta\omega^{ab,\left(  k-2\right)  }$ we find%
\begin{equation}
\delta_{\omega}S=\frac{2}{l^{2}}\alpha_{2k}\int\epsilon_{abcd}R^{a}e^{b}%
\delta\omega^{cd,\left(  k-2\right)  },
\end{equation}
with $k=2,\dots,m$; $R^{a}=R^{a,\left(  1\right)  }$ and $e^{a}=e^{a,\left(
1\right)  }$. \ One can see that the variation vanishes for arbitrary
$\delta\omega^{ab,\left(  k-2\right)  }$ if $R^{a}=0.$

Nevertheless it is possible to modify $\delta\omega^{ab,\left(  k-2\right)  }$
by adding an extra piece such that the variation of the action $\left(
\sim\alpha_{2k}\right)  $ can be written as%
\begin{equation}
\delta S=-\frac{4}{l^{2}}\alpha_{2k}\int R^{a}\left(  \bar{\Psi}\gamma
_{a}\gamma_{5}\epsilon^{\left(  k-1\right)  }-\frac{1}{2}\epsilon_{abcd}%
e^{b}\delta_{extra}\omega^{cd,\left(  k-2\right)  }\right)  .
\end{equation}
Thus the transformation of the $\omega^{ab,\left(  k-2\right)  }$ field
leaving the term proportional to $\alpha_{2k}$ invariant is%
\[
\delta_{extra}\omega^{ab,\left(  k-2\right)  }=2\epsilon^{abcd}\left(
\bar{\Psi}_{ec}\gamma_{d}\gamma_{5}\epsilon^{\left(  k-1\right)  }+\bar{\Psi
}_{de}\gamma_{c}\gamma_{5}\epsilon^{\left(  k-1\right)  }-\bar{\Psi}%
_{cd}\gamma_{e}\gamma_{5}\epsilon^{\left(  k-1\right)  }\right)  e^{e},
\]
with $\bar{\Psi}=\bar{\Psi}_{ab}e^{a}e^{b}$. \

On the other hand, it is important to note that the term proportional to
$\alpha_{2k}$ is truly invariant under gauge supersymmetry transformations
associated to $Q_{\left(  q\right)  }$, with $q\geq k$.

Clearly for $m=2$ in $s\mathcal{M}_{m+2}$ we recover the results presented in
the previous section.

\subsection{Pure supergravity from the minimal Maxwell algebra type
$s\mathcal{M}_{m+2}$}

\qquad Since we are interested in obtaining the Einstein-Hilbert and the
Rarita-Schwinger terms, we shall consider only the piece proportional to
$\alpha_{4}$. \ Then the action is written with the following choice for the
non-vanishing components of an invariant tensor%
\begin{align}
\left\langle J_{ab,\left(  0\right)  }J_{cd,\left(  4\right)  }\right\rangle
_{s\mathcal{M}_{m+2}}  &  =\alpha_{4}\left\langle J_{ab}J_{cd}\right\rangle
,\\
\left\langle J_{ab,\left(  2\right)  }J_{cd,\left(  2\right)  }\right\rangle
_{s\mathcal{M}_{m+2}}  &  =\alpha_{4}\left\langle J_{ab}J_{cd}\right\rangle
,\\
\left\langle Q_{\alpha,\left(  1\right)  }Q_{\beta,\left(  3\right)
}\right\rangle _{s\mathcal{M}_{m+2}}  &  =\alpha_{4}\left\langle Q_{\alpha
}Q_{\beta}\right\rangle ,
\end{align}
which can be written as%
\begin{align}
\left\langle J_{ab,\left(  0\right)  }J_{cd,\left(  4\right)  }\right\rangle
_{s\mathcal{M}_{m+2}}  &  =\alpha_{4}\epsilon_{abcd},\\
\left\langle J_{ab,\left(  2\right)  }J_{cd,\left(  2\right)  }\right\rangle
_{s\mathcal{M}_{m+2}}  &  =\alpha_{4}\epsilon_{abcd},\\
\left\langle Q_{\alpha,\left(  1\right)  }Q_{\beta,\left(  3\right)
}\right\rangle _{s\mathcal{M}_{m+2}}  &  =2\alpha_{4}\left(  \gamma
_{5}\right)  _{\alpha\beta}.
\end{align}
Then we only need the two-form curvatures asociated to $J_{ab,\left(
0\right)  }$, $J_{ab,\left(  2\right)  }$, $J_{\left(  ab,4\right)  }$,
\thinspace$Q_{\alpha,\left(  1\right)  }$ and $Q_{\alpha,\left(  3\right)  }$
which can be derived from $\left(  \ref{curvgen01}\right)  -\left(
\ref{curvgen3}\right)  $

Considering the different non-vanishing components of the invariant tensor and
the respective two-form curvatures we obtain the following action for the
$S$-expanded superalgebra%
\begin{equation}
S=2\alpha_{4}\int\left(  \frac{1}{2}\epsilon_{abcd}\mathcal{R}^{ab,\left(
0\right)  }\mathcal{R}^{cd,\left(  4\right)  }+\frac{1}{4}\epsilon
_{abcd}\mathcal{R}^{ab,\left(  2\right)  }\mathcal{R}^{cd,\left(  2\right)
}+\frac{4}{l}\bar{\Psi}^{\left(  3\right)  }\wedge\gamma_{5}\Psi^{\left(
1\right)  }\right)  ,
\end{equation}
which can be written explicitly as follows%
\begin{align}
S  &  =\alpha_{4}\int\epsilon_{abcd}\frac{1}{l^{2}}\left(  \mathcal{R}%
^{ab,\left(  0\right)  }e^{c,\left(  2\right)  }e^{d,\left(  2\right)  }%
+4\bar{\psi}^{\left(  1\right)  }e^{a,\left(  2\right)  }\gamma_{a}\gamma
_{5}D_{\omega}\psi^{\left(  1\right)  }\right) \nonumber\\
&  +d\left(  \epsilon_{abcd}\left(  \mathcal{R}^{ab,\left(  0\right)  }%
\omega^{ab,\left(  4\right)  }+\frac{1}{2}D_{\omega}\omega^{ab,\left(
2\right)  }\omega^{cd,\left(  2\right)  }\right)  \right. \nonumber\\
&  \left.  +\frac{8}{l}D_{\omega}\bar{\psi}^{\left(  1\right)  }\gamma_{5}%
\psi^{\left(  3\right)  }+\frac{1}{l}\bar{\psi}^{\left(  1\right)  }%
\omega^{ab,\left(  2\right)  }\gamma_{ab}\gamma_{5}\psi^{\left(  1\right)
}\right)
\end{align}
Here we have used the gravitino Bianchi identity $D\Psi^{\left(  1\right)
}=\frac{1}{4}R^{ab}\gamma_{ab}\Psi^{\left(  1\right)  }$ and the matrix gamma
identity $\left(  \ref{gamma}\right)  $ to show that
\begin{align*}
\epsilon_{abcd}\mathcal{R}^{ab,\left(  0\right)  }\bar{\psi}^{\left(
3\right)  }\gamma^{cd}\psi^{\left(  1\right)  }+8D\bar{\psi}^{\left(
1\right)  }\gamma_{5}D_{\omega}\psi^{\left(  3\right)  }  &  =D\left(
8D_{\omega}\bar{\psi}^{\left(  1\right)  }\gamma_{5}\psi^{\left(  3\right)
}\right) \\
\frac{1}{2}\epsilon_{abcd}D\omega^{ab,\left(  2\right)  }\bar{\psi}^{\left(
1\right)  }\gamma^{cd}\psi^{\left(  1\right)  }+2\bar{\psi}^{\left(  1\right)
}\omega^{ab,\left(  2\right)  }\gamma_{ab}\gamma_{5}D\psi^{\left(  1\right)
}  &  =D\left(  \bar{\psi}^{\left(  1\right)  }\omega^{ab,\left(  2\right)
}\gamma_{ab}\gamma_{5}\psi^{\left(  1\right)  }\right)  .
\end{align*}
Then using the following identification%
\begin{align*}
\omega^{ab,\left(  0\right)  }  &  =\omega^{ab},\text{ \ \ }\omega^{ab,\left(
2\right)  }=\tilde{k}^{ab},\\
\omega^{ab,\left(  4\right)  }  &  =k^{ab},\text{ \ \ }e^{a,\left(  2\right)
}=e^{a},\\
\mathcal{R}^{ab,\left(  0\right)  }  &  =R^{ab},\text{ \ \ }\psi^{\left(
1\right)  }=\psi,\\
\psi^{\left(  3\right)  }  &  =\xi,
\end{align*}
the action is given by%
\begin{align}
S  &  =\alpha_{4}\int\epsilon_{abcd}\frac{1}{l^{2}}\left(  R^{ab}e^{c}%
e^{d}+4\bar{\psi}e^{a}\gamma_{a}\gamma_{5}D_{\omega}\psi\right) \nonumber\\
&  +d\left(  \epsilon_{abcd}\left(  R^{ab}k^{cd}+\frac{1}{2}D_{\omega}%
\tilde{k}^{ab}\tilde{k}^{cd}\right)  +\frac{8}{l}\bar{\xi}\gamma_{5}D_{\omega
}\psi+\frac{1}{l}\bar{\psi}\tilde{k}^{ab}\gamma_{ab}\gamma_{5}\psi\right)
\end{align}

Here we can see that the action proportional to $\alpha_{4}$ contains the
Einstein-Hilbert term $\epsilon_{abcd}R^{ab}e^{c}e^{d}$, the Rarita-Schwinger
lagrangian $4\bar{\psi}e^{a}\gamma_{a}\gamma_{5}D\psi$ and a boundary term
involving the new fields $k_{ab},$ $\tilde{k}_{ab},~\xi$ and the original ones.

Unlike the Mac Dowell-Mansouri lagrangian for $\mathfrak{osp}\left(
4|1\right)  $ superalgebra the supersymmetric cosmological constant does not
appear explicitly in this action. \ This is due to the $S$-expansion procedure
since if we want to obtain the supersymmetric cosmological constant%
\[
\frac{1}{2l^{4}}e^{a}e^{b}e^{c}e^{d}+\frac{1}{l^{3}}\bar{\psi}\gamma^{ab}\psi
e^{c}e^{d}%
\]
in the action, it should be necessary to consider the components $\left\langle
J_{ab,\left(  4\right)  }J_{cd,\left(  4\right)  }\right\rangle $ and
$\left\langle J_{ab,\left(  2\right)  }J_{cd,\left(  4\right)  }\right\rangle
$ which are proportional to $\alpha_{8}$ and $\alpha_{6}$, respectively.\

Independently of the numbers of new generators of the Maxwell superalgebra,
the new Maxwell fields do not contribute to the dynamics of the term
proportional to $\alpha_{4}$. \ In this way, we have shown that $N=1$, $D=4$
pure supergravity can be obtained as a Mac Dowell-Mansouri like action for the
minimal Maxwell superalgebras $s\mathcal{M}_{m+2}$ $\left(  \text{with
}m>1\right)  $.
\begin{equation}
S=\alpha_{4}\int\frac{1}{l^{2}}\left[  \epsilon_{abcd}R^{ab}e^{c}e^{d}%
+4\bar{\psi}e^{a}\gamma_{a}\gamma_{5}D\psi\right]  +\text{ boundary terms.}%
\end{equation}

It is important to note that the case $m=1$ corresponds to $D=4$ Poincar\'{e}
superalgebra $s\mathcal{P}=\left\{  J_{ab},P_{a},Q_{\alpha}\right\}  $ as
mencioned in Ref. \cite{15}. \ However it is not possible to derive the pure
supergravity action from a Mac Dowell-Mansouri like action for this
superalgebra since it is not possible to obtain the Eintein-Hilbert term from
$\left\langle J_{ab}J_{cd}\right\rangle $ for $s\mathcal{P}$.

\section{Comments and possible developments}

\qquad In the present work we have derived the minimal $D=4$ supergravity
action from the minimal Maxwell superalgebra $s\mathcal{M}_{4}$. \ For this
purpose we have applied the abelian semigroup expansion procedure to the
$\mathfrak{osp}\left(  4|1\right)  $ superalgebra allowing us to build a Mac
Dowell-Mansouri like action. \ Interestingly, the action obtained describes
pure supergravity in four dimensions. \ This result can be seen as a
supersymmetric generalization of Ref. \cite{12} in which four-dimensional
General Relativity is derived from Maxwell algebra as a Born-Infeld like action.

We have also obtained the $D=4$ supergravity action from the minimal Maxwell
superalgebra type $s\mathcal{M}_{m+2}$ using bigger semigroups. \ These
superalgebras enlarge the pure supergravity action adding new terms containing
new Maxwell symmetries. \ In particular, the action found in Ref.\ \cite{18}
can be obtained when the simplest Maxwell superalgebra is considered.\ The
invariance of the actions under the new supersymmetry transformations has been
analized in detail.

Our results provide another example showing that the $S$-expansion procedure
is not only a useful method to derive new lie superalgebras but it is a
powerful and simple tool in order to construct a supergravity action for an
$S$-expanded superalgebra. \ Moreover the invariance of the pure supergravity
action under new supersymmetry transformations could not be guessed trivially.

A future work could be consider the $N$-extended Maxwell superalgebras and the
construction of $N$-extended supergravities in a very similar way to the one
shown here. It would be also interesting to build lagrangians in odd
dimensions using the Chern-Simons formalism and the $S$-expansion method. \ It
seems that it should be possible to recover standard odd-dimensional
supergravity from the Maxwell superalgebras [work in progress].

\section*{Acknowledgements \textbf{\ }}

\qquad The authors wish to thank L. Andrianopoli, R. D'Auria and M. Trigiante
for enlightening discussions and their\ kind hospitality at Dipartimento
Scienza Applicata e Tecnologia of Politecnico di Torino, where this work was
done. \ We are also grateful to P. Salgado for introducing the topics covered
in the present work.\ The autors were supported by grants from the
Comisi\'{o}n Nacional de Investigaci\'{o}n Cient\'{\i}fica y Tecnol\'{o}gica
(CONICYT) and from the Universidad de Concepci\'{o}n, Chile. \ This work was
supported in part by FONDECYT Grants N$%
{{}^\circ}%
$ 1130653

\end{document}